# Distributed Link Scheduling with Constant Overhead

Sujay Sanghavi
LIDS, MIT
sanghavi@mit.edu

Loc Bui
CSL, UIUC
locbui@uiuc.edu

R. Srikant
CSL, UIUC
rsrikant@uiuc.edu

## ABSTRACT

This paper proposes a new class of simple, distributed algorithms for scheduling in wireless networks. The algorithms generate new schedules in a distributed manner via simple local changes to existing schedules. The class is parameterized by integers $k \geq 1$. We show that algorithm $k$ of our class achieves $k/(k+2)$ of the capacity region, for every $k \geq 1$.

The algorithms have small and constant worst-case overheads: in particular, algorithm $k$ generates a new schedule using *(a)* time less than $4k+2$ round-trip times between neighboring nodes in the network, and *(b)* at most three control transmissions by any given node, for any $k$. The control signals are explicitly specified, and face the same interference effects as normal data transmissions.

Our class of distributed wireless scheduling algorithms are the first ones guaranteed to achieve any fixed fraction of the capacity region while using small and constant overheads that do not scale with network size. The parameter $k$ explicitly captures the tradeoff between control overhead and scheduler throughput performance and provides a tuning knob protocol designers can use to harness this trade-off in practice.

## 1. INTRODUCTION

This paper presents novel distributed algorithms for scheduling transmissions in wireless networks. The algorithms represent the first instance in which any arbitrary fraction of the capacity region can be achieved with constant overhead. In addition, our algorithms are very simple. We now motivate our work and summarize our contributions.

The task of wireless scheduling is challenging due to the simultaneous presence of two characteristics: interference between transmissions, and the need for practical distributed implementation. In any given wireless network, interference effects result in a fundamental upper limit on the data rates that any scheduling algorithm – distributed or otherwise – can hope to achieve. This fundamental limit, or capacity region, serves as a benchmark against which the performance of various distributed scheduling algorithms can be compared.

In practice, the need for distributed implementation invariably leads to an overhead, as the same time, power and bandwidth resources that could have been used for data transmission have to, instead, be wasted on control signals in an effort to combat interference. Most of the previously proposed scheduling algorithms, which we survey and compare in Section 2, manage to achieve the capacity region (or approximations thereof) using algorithms in which the time and communication overheads *grow with network size*.

Most of these algorithms attempt to maximize scheduling performance while ignoring (i.e. not explicitly accounting for) the control overheads. This is clearly a problem from the perspective of modelling wireless resource usage. In particular, it may be the case (especially for large networks) that after using a large – and unaccounted for – portion of resources for control signalling, the algorithms perform well with regards to the benchmark in the *remaining* portion used for data transmission.

Scheduling algorithms with growing overheads exacerbate this problem, to the extent that it is in general not immediately clear what relation the claimed scheduling performance[1] has to the actual efficiency of overall resource utilization in general networks. In recognition of this fact, the papers make (sometimes convincing, but heuristic) arguments as to why their algorithms have manageable overheads.

In our paper we adopt a more principled approach of taking the overheads into account a priori in the performance evaluation. This fact, along with the fact that our algorithms have constant overheads, implies that for our algorithms it is clear *(i)* how efficiently the overall wireless resources are utilized, *(ii)* how we can tradeoff between scheduling performance and control overhead, and *(iii)* how a system designer can choose her operating point on this tradeoff. These three aspects are elaborated on below, after a brief description of our results.

As an aside, note that the modelling probem outlined above is not present in the (algorithmically closely related) task of switch scheduling. This is because in switches the computation resources used in the overheads are *separate from* the constrained resources that need to be efficiently scheduled – namely, the crossbar switch. Growing overheads thus do not need to figure in the accounting of the utilization

[1] which is typically compared to the benchmark "capacity region" or "100% throughput region"

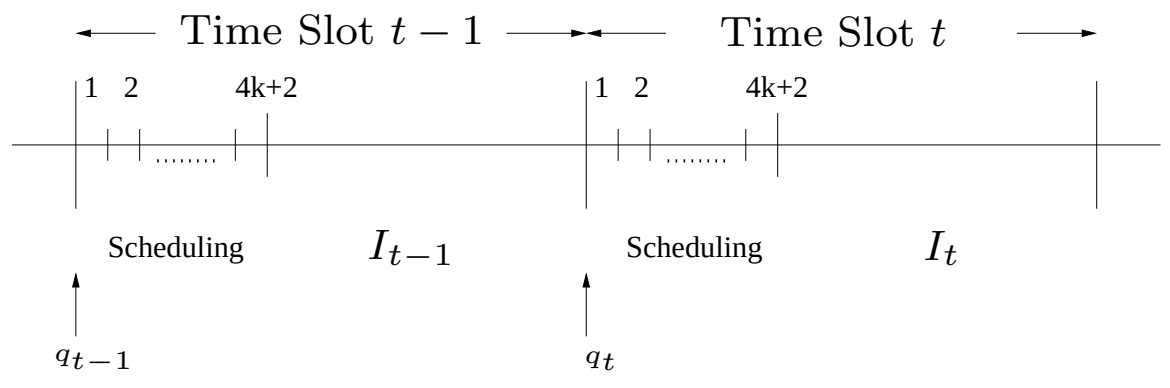

**Figure 1: Each scheduling cycle is divided into a control part and a data part. The control part consists of $4k+2$ phases – each phase being the length of a round-trip between neighbors – after which a new set of active links will be decided. This new set is active for the data part of the cycle. The whole process is repeated in the next cycle with updated queues. $t$ counts the cycle number.**

of the constrained resource. However, in wireless networks the overheads use the *same* resource as the one that has to be efficiently scheduled, and hence growing overheads *are* a problem. The need for fast algorithms in switches is driven primarily by practical hardware limitations.

Growing overheads are also undesirable from a wireless protocol implementation viewpoint. Besides possible resource wastage, growing overheads have the potential to introduce dependencies on network size into MAC scheduling, complicating protocol design. We are not the first to recognize the need for protocols with constant overheads: some recent pieces of work [1, 2, 3] also propose constant-overhead algorithms. Their results and approaches are summarized and compared to ours in Section 2. These existing constant-overhead algorithms can guarantee at most half of the capacity region (in the portion of resources dedicated to data transmission), and essentially involve doing enhanced contention resolution as a way to approximate maximal matching in constant time. As opposed to these protocols, our algorithms can capture any desired fraction of the capacity region (in the data transmission part), and do not attempt enhanced contention resolution. Thus our work differs from the existing work in methodology, and goes beyond in terms of performance.

In this paper, we present a class of simple distributed algorithms for scheduling in the "node exclusive" or "primary" interference model. In this model any node in the network can communicate with at most one other node at any time. This is an important model with a rich history of dedicated work, which we survey in Section 2.

In our algorithms bandwidth is assumed to be fixed and time is divided into scheduling cycles, with a new schedule generated by the algorithm in every cycle. In our paper, as in others in this area, the length of a cycle is left to the protocol designer. Our algorithms partition each cycle into two parts: a scheduling (control signalling) part and a service (data-transmission) part. Figure 1 depicts this partition.

The algorithm in our class corresponding to parameter value $k$ requires that the length of the scheduling part be $4k+2$ round-trip times, where one *round-trip time* is the amount of time required for a node to make a very basic two-way handshake with a neighboring node. This algorithm is then guaranteed to achieve a fraction $\left(\frac{k}{k+2}\right)$ of the capacity region *during the data part* of the cycle, and for any network. Also, in any scheduling cycle, the algorithm requires at most three control signal transmissions, for any $k$. Also, the size of each control transmission is fixed a priori and is independent of $k$ or network size. Finally, each node may have to perform at most one computation, which is just taking the difference of two integers given to it.

The partitioning of the scheduling cycle thus explicitly captures the wastage in control signalling: the fraction of resources wasted is the ratio of the length of the control part to the length of the overall cycle.

A larger value of the parameter $k$ requires a longer absolute length of the control part, and in return guarantees better performance in the data part, as explicitly detailed above. Thus $k$ captures the overhead-performance tradeoff. Our class thus provides the protocol designer a tunable knob with which to optimize performance, with respect to other system considerations. One such consideration that may have a direct bearing on the appropriate choice of $k$ is the length of the scheduling cycle. If long cycles are determined to be feasible – where "long" is as compared to the round-trip time – it may make sense to choose a protocol with larger $k$. Conversely, short scheduling cycles may favor a small-$k$ implementation. The choice of the parameter $k$ may depend on network characteristics like mobility and arrival statistics; however, *it does not depend on network size.*

In the rest of the paper we concentrate on two objectives: specifying a scheduling algorithm that runs with the above-mentioned control overheads, and showing that this algorithm achieves the claimed capacity region during the data part of the cycle. Thus, in the rest of the paper, the term "capacity region" has to be interpreted as the capacity of the data portion of the overall cycle. This will be compared to the "100% throughput region" as it applies to the data part.

The rest of the paper is organized as follows. In Section 2 we survey some of the existing literature most relevant to our paper, and compare our results. Section 3 lays out the formal system model for scheduling in the presence of primary interference. Section 4 presents our parameterized class of algorithms, along with examples, illustrations and discussion. Section 5 proves the performance of our algorithm for arbitrary networks. In Section 6 we investigate the performance of our algorithm for a simple grid network. We conclude with a discussion in Section 7.

## 2. BACKGROUND AND EXISTING WORK

Scheduling in the presence of interference constraints is a central problem in communication networks. In this summary, we will mainly concentrate on the work involving primary interference constraints, also known as the "node-exclusive" model in wireless networks. Primary interference constraints arise both in wireless networks and input-queued crossbar switches in Internet routers, and the results of the papers listed below are often of interest in both applications. In the following, "complexity" refers to the number of operations/amount of time that has to be spent every time a new schedule has to be found.

Hajek and Sasaki [4] introduced the primary interference model, which they studied in the wireless context and for fixed given arrivals. Tassiulas and Ephremides [5] were the first to consider stochastic arrivals in general interference models, of which primary interference is a special case. They characterized the maximum attainable capacity region, and also presented a centralized algorithm guaranteed to achieve it. In the case of primary interference, this algorithm boils down to finding maximum weight matchings (with queue lengths being weights). This algorithm thus has $\Omega(n^2)$ complexity. McKeown et. al. [6] also showed the same result for switches.

The need for speedy implementation and low overhead spurred the development of algorithms with lower complexity (but possibly higher delays). Tassiulas [7] studied randomized centralized algorithms that achieve the capacity region with $O(n)$ complexity. This algorithm samples a new candidate matching uniformly from the set of all matchings, and switches schedules to this new sample if and only if it represents a larger weight. For the case of switches, this algorithm was de-randomized by Giaccone et. al. [8]. A distributed implementation of [7] for wireless networks was proposed by Modiano et. al. [9], with the weight comparison between matchings being done via an averaging mechanism.

Weller and Hajek [10] showed that any algorithm that uses a maximal matching in every time slot can achieve half the capacity region. They showed this result for deterministically upper-constrained traffic. Dai and Prabhakar [11] showed the same performance holds for stochastic packet arrivals as well. Lin and Shroff [12] extended this result to the case of flow arrivals and departures.

Recently, distributed algorithms achieving the entire capacity region have been proposed, see e.g. [9, 13, 14, 15]. This guarantee of course refers to the scheduling efficiency with regards to data transmission, since there papers do not account for resources used in overheads. Also, these protocols have overheads that grow with network size.

All of the above algorithms involve complexities that grow with network size. In some more recent work scheduling algorithms with constant overheads have been proposed. Lin and Rasool [1] showed that close to 1/3 of the capacity region can be achieved with $O(1)$, i.e. constant, overhead. Gupta et.al. [2] and Joo and Shroff [3] build on this result to achieve close to 1/2 the capacity region with constant overheads. These algorithms attempt to generate (approximately) maximal matchings in every time slot using local contention algorithms that terminate in $O(1)$ time. Our approach in this paper is thus different from these papers, as we do not attempt to resolve contention.

All the above papers consider single-hop traffic. Some more recent developments have pushed in the directions of multi-hop traffic and more general interference constraints: see for example [16, 17, 18].

# 3. PRELIMINARIES

We now describe the (standard) model for scheduling in the presence of primary interference, with the implicit understanding that the capacities and bounds refer to what is achievable in the service part of a time slot. Consider a wireless network modeled by a graph $G = (V, E)$, where $V$ is the set of nodes and $E$ is the set of links. We assume that the time is slotted, denoted by $t$. Nodes communicate data to other nodes in the form of packets, whose size is normalized so that each packet can be communicated in one time slot. All traffic is assumed to be single-hop. Let $A_t(e)$ denote the number of packets arriving at time $t$ for transmission over link $e$, and $A_t$ be the vector of all arrivals at time $t$. The packets in $A_t$ can be communicated at time slot $t$ or later. The arrival process $A_t$ is assumed to be independent and identically distributed across time[2], with average arrival rate vector $a = E[A_t]$ and bounded second moment $E[A_t' A_t] < \infty$. The arriving packets are stored in queues, of which there is one for each link. Let $q_t(e)$ be the queue length associated with link $e$ at time $t$, and $q_t$ be the vector of queue lengths at time $t$. We assume there is no a priori upper bound on the maximum queue size, so there are delays but no packet drops.

Links can be *active* or *inactive*. Each active link can transmit one packet in its queue. In this paper we work with the *node-exclusive spectrum sharing* model of primary interference in wireless networks, which requires that any node communicate with at most one other node in any time slot. This means the set of simultaneously active links is constrained to be a *matching* in $G$. We will use the binary vector $I_t$ of length $|E|$ to denote the set of active links at time $t$, with the convention that $I_t(e) = 1$ if and only if link $e$ is active and has a positive queue at time $t$. The queue lengths thus evolve according to

$$q_{t+1} \;=\; q_t + A_{t+1} - I_t$$

Let $\mathcal{I}$ be the set of all feasible matching vectors $I$ in $G$. The *capacity region* $\mathcal{C}$ of the network is the strict convex closure of all matchings: $a \in \mathcal{C}$ if and only if there exist non-negative numbers $\lambda_1, \ldots, \lambda_{|\mathcal{I}|}$ such that

$$a \;=\; \sum_m \lambda_m I^{(m)} \quad \text{and} \quad \sum_m \lambda_m \;<\; 1$$

$\mathcal{C}$ has also been referred to in the literature as the "stability region" and the "100% throughput region". It is well known [5] that any rate vector $a \notin \mathcal{C}$ will lead to some queues in the network being unstable. The task of a link scheduling algorithm is to determine which links are to be activated at any given time. Any algorithm which guarantees stable queues for any $a \in C$ is said to "achieve $\mathcal{C}$".

In this paper we will be interested in algorithms that achieve a fraction $0 \le \beta \le 1$ of the capacity region. The fraction $\beta\mathcal{C}$ of the capacity region is the set of all arrival rates $a$ for which there exist non-negative numbers $\lambda_1, \ldots, \lambda_{|\mathcal{I}|}$ such that

$$a \;=\; \sum_m \lambda_m I^{(m)} \quad \text{and} \quad \sum_m \lambda_m \;<\; \beta \tag{1}$$

An algorithm is said to achieve $\beta\mathcal{C}$ if it can ensure queues are stable for any $a \in \beta\mathcal{C}$.

In our algorithms, the decision for which links become active at time $t$ is based on the pair $(q_t, I_{t-1})$ of current queue lengths and the last available schedule of active links. Consider the algorithm corresponding to parameter $k$. Figure 1 summarizes the service structure for this algorithm. Each time slot is divided into a scheduling part and a service part. The scheduling part consists of $4k + 2$ *phases*, during which the set of active links in $I_t$ will be decided based on $(q_t, I_{t-1})$. This new set of links remains active for the service part of time slot $t$. The whole process is repeated in the next time slot with updated queues. The length of a phase is one round-trip time between adjacent nodes in $G$.

# 4. THE ALGORITHM

[2]It may be correlated across links. Time correlation can also be allowed, but at the expense of more complicated proofs.

In this section we present our algorithm for determining the new schedule $I_t$ from $(q_t, I_{t-1})$. To do so, we will need a few simple definitions. In the following we will abuse notation by letting $I$ denote matchings as well as the associated $\{0,1\}$ valued vector of length $|E|$: $I(e) = 1$ if and only if link $e$ is in the matching $I$.

Recall that in $I$ no two adjacent links can be active. An *augmentation* $A$ of a matching $I$ is a path or cycle in which every alternate link is in $I$, with the property that if all links in $A \cap I$ are removed from $I$ and all links in $A - I$ are added[3] to $I$, then the resulting set of links will remain a matching in $G$. This process of changing $I$ using $A$ is called *augmenting* $I$ with $A$, and the resulting augmented matching is denoted by $I \oplus A$.

The *size* of augmentation $A$ is the number $|A - I|$ of links in $A$ that are not also in $I$. Note that any augmentation of size $k$ will have at most $2k+1$ links in it.

Two augmentations $A_1$ and $A_2$ are *disjoint* if no two links in $A_1 - I$ and $A_2 - I$ are adjacent. This implies that $I$ can be simultaneously augmented by $A_1$ and $A_2$ and still be a valid matching[4]. $\mathcal{A}$ is a set of disjoint augmentations if every pair in $\mathcal{A}$ is disjoint. Clearly, if $\mathcal{A}$ is a set of disjoint augmentations then $I \oplus \mathcal{A}$ will be a matching in $G$.[5]

Finally, for any time $t$, the *gain* of an augmentation $A$ to $I_{t-1}$ – the matching used in the previous time slot – is defined as

$$gain_t(A) \quad := \quad \sum_{e \in A - I_{t-1}} q_t(e) - \sum_{e \in A \cap I_{t-1}} q_t(e) \qquad (2)$$

If the queue lengths $q_t(e)$ are considered to be the weights on links, then $gain_t(A)$ represents the change in the weight of $I_{t-1}$ if it is augmented by $A$. Similarly, the gain of a set $\mathcal{A}$ of disjoint augmentations is the sum of the gains each augmentation in that set.

Our algorithm obtains $I_t$ by augmenting $I_{t-1}$ with a set of disjoint augmentations of size at most $k$ and positive gain. The process is illustrated in the example below:

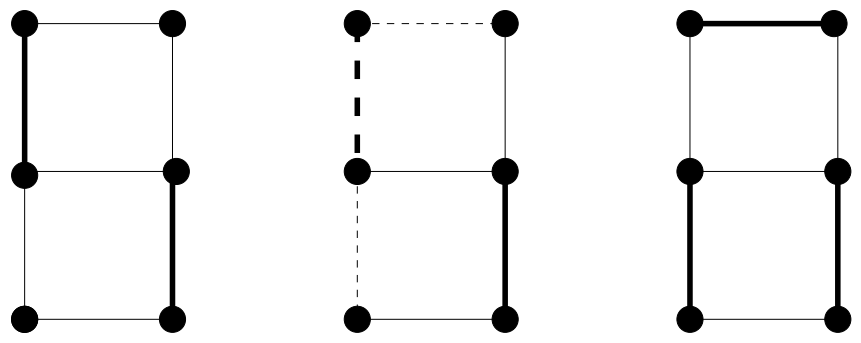

The bold lines in the left-most figure indicate links in $I_{t-1}$. The dotted lines in the central figure indicate the links in augmentation $A$. The bold-dotted lines are in $A \cap I_{t-1}$, and the thin-dotted lines are in $A - I_{t-1}$. $I_t$ is obtained by augmenting $I_{t-1}$ with $A$. The bold lines in the last figure are the links in $I_t$. The idea of using fixed-length augmentations to obtain aproximations to maximum-weight matching has been used previously (see e.g. Pettie and Sanders [19]) in a different, pure graph-theoretic context to find an approximation to maximum-weight matching with linear complexity.

[3]Here $A - I$ denotes the set of links in $A$ but not in $I$.

[4]Note that a link in $I$ can be part of both $A_1$ and $A_2$ even if they are disjoint.

[5]$I \oplus \mathcal{A}$ is $I$ augmented by every $A \in \mathcal{A}$. The fact that the augmentations are disjoint means that this can be done in any order.

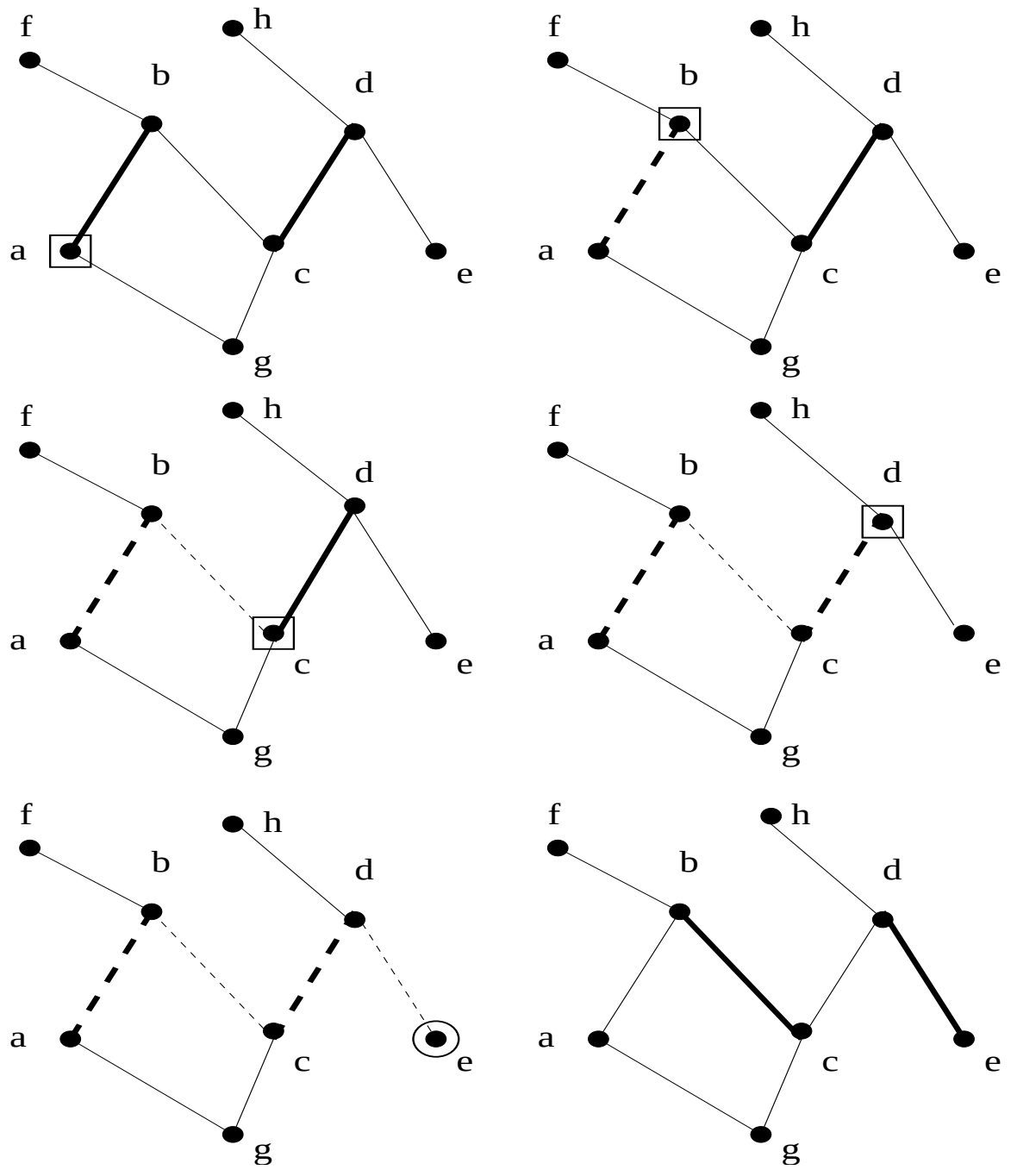

**Figure 2: Augmentation building in our algorithm (Example 1). The □ indicates active nodes. Dashed lines are links are in $A$. Bold lines in the first 5 graphs are links in $I_{t-1}$, and in the last graph are links in $I_{t-1} \oplus A$. ◯ indicates terminus.**

In our algorithm, augmentations are built in a random distributed fashion. We now present a simplified example to aid in the visualization of how our algorithm makes one augmentation.

**Example 1: Augmentation Building in the Absence of Contention**

Consider Figure 2, which depicts our algorithm operating in phases on a simple graph. The bold links in the top left graph of Figure 2 are the links of $I_{t-1}$.

In our algorithm, an augmentation begins at a *seed* and ends at a *terminus*. This seed is *active* in phase 1. When a node is active, it tries to elongate its augmentation $A$ by adding links as follows:

1. if the node has a link in $I_{t-1}$ that is not already in $A$, it adds $I_{t-1}$ to $A$.

2. else, it adds a random new link to $A$.

Every time a link is added, for the next phase the currently active node becomes inactive, and the new end-node becomes active instead.

As seen in the first figure, node $a$ is the only seed and is active in phase 1. It adds link $(a,b)$ to its (currently empty) augmentation in phase 1, since $(a,b) \in I_{t-1}$ and it is not currently in $a$'s augmentation. For phase 2, $a$ is inactive and $b$ is active.

Node $b$ can choose to add any one of the links $(b,c)$ or $(b,f)$. Say it chooses $(b,c)$. So, for phase 3, node $c$ is active and $b$ is inactive.

Node $c$ adds the link $(c,d)$, because it is a new link in $I_{t-1}$. So in phase 4 $d$ becomes active. $d$ picks randomly from links $(d,e)$ and $(d,h)$. Say it picks $(d,e)$.

Node $e$ would have become active as a result, but it has no further links to add to the augmentation. So it instead becomes the *terminus* of the augmentation.

Terminus $e$ then evaluates the gain

$$gain_t(A) \; = \; q_t(b,c) + q_t(d,e) - q_t(a,b) - q_t(c,d)$$

using net queue length information that has been passed on along $a, b, c, d, e$ during the building of the augmentation. In our example, it finds that $gain_t(A) > 0$ and decides to switch. This decision is then passed on back along the links $(e, d)$, $(d, c)$,$(c, b)$ and $(b, a)$ over the next 4 phases. Then, the links in $A$ are switched to obtain the final graph above, where bold links are the ones in $I_{t-1} \oplus A$. ■

The above algorithm illustrates the simple basic idea underlying our algorithm, namely: *(i)* randomly seed and grow disjoint augmentations, and *(ii)* switch all the augmentations that have positive gain.

The above example illustrated the augmentation building procedure in an idealized network where it was the only augmentation. In an actual wireless network however, augmentations are seeded at random. This means that there will be multiple augmentations, which will have to contend for access to links while ensuring that they remain consistent (i.e. valid augmentations) and disjoint.

We now succinctly describe the algorithm that makes $I_t$ from $(q_t, I_{t-1})$. A more realistic, and more detailed, example and a brief discussion follows the description.

In our algorithm each augmentation builds up in phases starting from a seed and ending in a terminus. For any phase, and node $v$, let $aug(v)$ denote the augmentation (if any) that $v$ is a part of in that phase. Also, the term "new link" for a node $v$ refers to any link $(u, v)$ that is not already in $aug(v)$. For any active $v$ except the seed, one of its links will be in $aug(v)$ and all the others will be new. Similarly, a "new neighbor" for $v$ is any node that shares with $v$ a link that is new for $v$.

We require the first link a seed adds be a link in $I_{t-1}$, if one is incident on the seed. Else it adds a random link. In each phase from 2 to $2k+1$, every augmentation alternately needs links in $I_{t-1}$ and links outside $I_{t-1}$.

## Algorithm Description

1. **Initialization:** Before phase 1,

   (a) each node randomly decides to be a *seed* with probability $p$.

   (b) Each seed chooses an *intended size* for its augmentation, uniformly from the set $\{1, \ldots, k\}$.

   The seeds are the *active* nodes in phase 1.

2. **Iteration:** Every seed $s$ that has some link $(s, r) \in I_{t-1}$ adds $(s, r)$ to $aug(s)$ and sends a REQ to the corresponding $r$ in phase 1. Every other seed sends a REQ along a random link. In each phase from 2 to $2k+1$, each active node $v$ tries to extend its $aug(v)$ as follows:

   (a) If $aug(v)$ needs a new link $I_{t-1}$, and if one such link $(v, u) \in I_{t-1}$ exists, then $(v, u)$ is added to $aug(v)$. Also, $v$ sends a REQ to $u$ along $(v, u)$. If no such link exists, $v$ becomes *terminus*.

   (b) If $aug(v)$ needs a new link outside $I_{t-1}$, but $size(aug(v))$ is already as intended, $v$ sends no REQ and becomes terminus. Otherwise, $v$ sends REQ to a random, uniformly chosen, new neighbor. If no new neighbor exists $v$ becomes terminus.

   The REQs above may face contentions. In any phase if active $v$ sends REQ to $w$ and

   (a) any another active $u$ also sends REQ to $w$ in that phase, a *collision* occurs. $w$ does not ACK, and $v$ becomes terminus.

   (b) If $w$ is a *used* node, i.e. it is already part of an augmentation, then it $w$ does not ACK and $v$ becomes terminus.

   (c) If $w$ is unused and there is no collision, then $w$ sends ACK to $v$. The link $(v, w)$ is added to $aug(v)$. $w$ will become active in the next phase and $v$ will become inactive.

   Every node that becomes terminus is inactive in subsequent phases.

3. **Termination:** After $2k+1$ phases, every terminus $w$ checks the following three conditions

   (a) if it is adjacent to its seed $v$,

   (b) if $aug(w)$ began and ended with links in $I_{t-1}$,

   (c) if $aug(w)$ has not reached its intended size.

   If all of above are true, link $(u, w)$ is added to $aug(w)$.

   Also, in either case, every terminus $w$ evaluates $gain_t(aug(w))$ and makes the decision of switching if and only if $gain_t(aug(w)) > 0$.

4. **Back-propagation and Switching:** Switching decision is relayed back in phases $2k+2$ to $4k+2$ from terminus to seed, along the path of each augmentation. These communications will be non-interfering. After phase $4k+2$, all nodes in augmentation implement decision.

## Discussion

Note that in our algorithm there is a small but crucial difference between links in $I_{t-1}$ and links outside $I_{t-1}$ with regards to the timing of link addition to an augmentation. Specifically, a link in $I_{t-1}$ is added *before* a REQ is sent (and irrespective of whether an ACK is received), while a link outside $I_{t-1}$ is added only after a REQ is sent *and* and ACK is received. This difference in timing ensures that augmentations are consistent, i.e. whenever a link $e \notin I_{t-1}$ is added to $A$, all the links in $I_{t-1}$ adjacent to $e$ are *also* added to $A$.

It is easy to see that our algorithm will ensure disjointness of augmentations. Consider the addition of a link $(v, u) \notin I_{t-1}$ to $aug(v)$ by an active node $v$ in some phase. This will only happen if $v$ sends a REQ to $u$ in that phase and $u$ responds with an ACK. Now, if $u$ is already a member of an augmentation by that phase (including the case of it being already a member of $aug(v)$ itself) then $u$ will not respond with an ACK. Also, if any other augmentation tries to add some other link $(w, u) \notin I_{t-1}$ at the same phase, then neither augmentation will be successful. Thus no two adjacent links outside $I_{t-1}$ will be part of augmentations. Thus all augmentations are consistent and disjoint. Note of course that any link $(z, u)$ that is in $I_{t-1}$ *can* be added to another augmentation. This however does not hurt the disjointness.

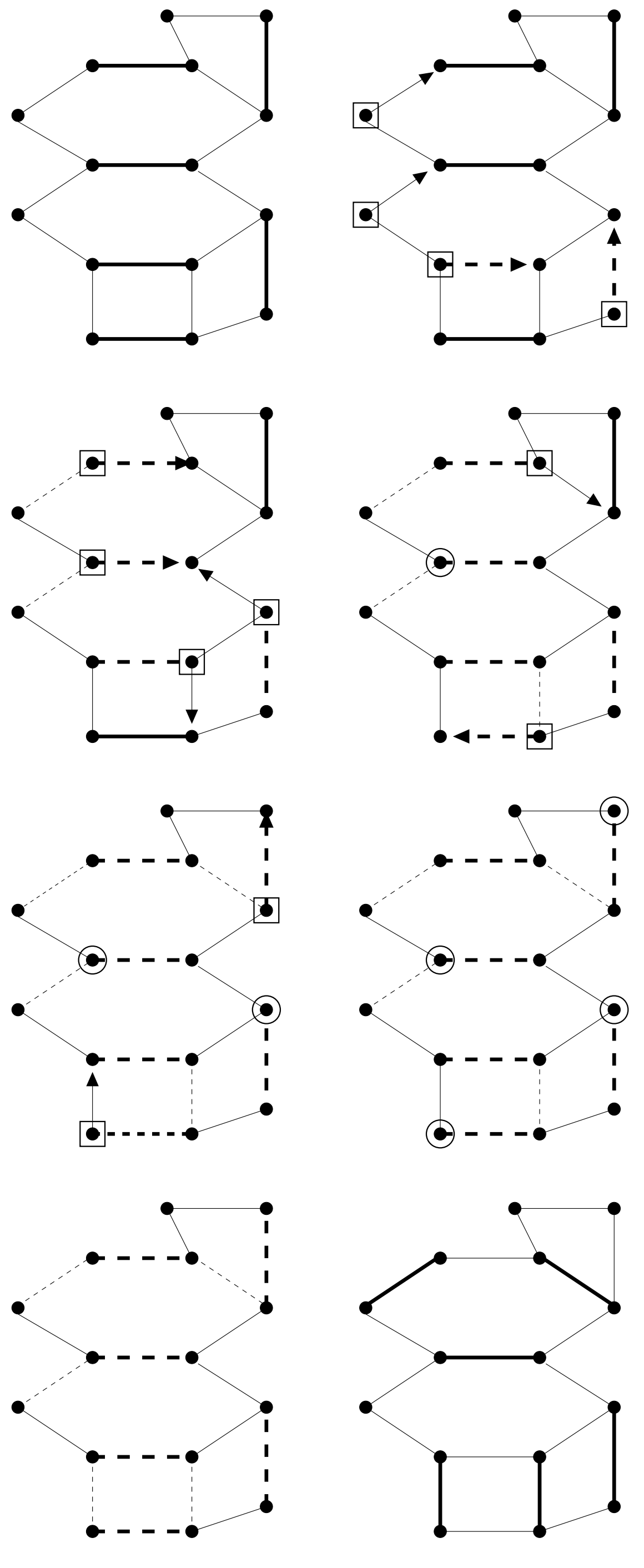

**Figure 3: The working of our algorithm in phases (Example 2). Bold lines in first seven graphs are links in $I_{t-1}$, in the last graph are links in $I_t$. Dashed lines are links in augmentations. □ denotes active nodes and ◯ denotes terminus nodes. Arrows depict the REQ signals in each phase.**

We now illustrate how our algorithm works by means of a slightly detailed example.

**Example 2: Working of the Algorithm**

Consider the network shown in Figure 3. The first graph shows the schedule $I_{t-1}$ of the previous time slot. The second graph shows the nodes active in phase 1, which are the seeds of the network. Assume $k = 2$ and also that the intended size chosen by every seed is also 2. Note that the active nodes that have a link in $I_{t-1}$ incident on them immediately include it in their augmentations, and also send the REQ along that link. Other active nodes do not include any links in their augmentations, and send REQs at random.

Since all REQs went to different targets, there were no collisions and so there are four new nodes active in phase 2. Again the newly active nodes that have a new link in $I_{t-1}$ incident on them immediately add it to their augmentation and send REQs along them. Other newly active nodes send REQs at random.

We see that two of the REQs sent in phase 2 will collide. This means that the corresponding transmitters of the collied REQs will become terminus of their augmentations, which then stop growing. The other two augmentations continue growing.

In phase 4, the active node at the bottom sends REQ to a used node (its own seed in this case). Hence, it does not receive an ACK, and becomes the terminus. The other active node in phase 4 adds its link without problems. However, the addition of another link to this augmentation would make it exceed its intended size. Thus after the addition the new node becomes terminus and does not further add links.

After the last phase 5, note that one of the augmentations (the bottom one) satisfies the conditions in step 4 of the algorithm, namely: it began and ended with links in $I_{t-1}$, its seed and terminus are joined by a link not in $I_{t-1}$ and the addition of this link does not violate the intended size. Thus, this link is added to get the final set of disjoint augmentations after phase 5 (=$2k + 1$).

These augmentations are switched depending on their gains. In our example, two of the four augmentations are switched and the other two are not. The switching decisions are relayed back in phases 6 to 10, after which the switching happens. The final graph depicts the resulting $I_t$. ■

The example illustrates the fact that without possible the last link addition in the "termination" part of the above algorithm, we would not be able to build augmentations that are (small) cycles.

Note that the information relayed on at each phase does not grow with $k$ or the network size: indeed, all that is needed is the net gain of the augmentation up until the current phase, and the identity of the first node (so that the termination part of the algorithm can be implemented).

The probability $p$ in the above algorithm is a parameter that can be chosen by the system designer. If $p$ is too high, there will be too many seeds in the network which will result in too much contention and not enough augmentations. if $p$ is too small, there will not be enough seeds to ensure a good enough set of augmentations.

## 5. ANALYSIS

In this section we will prove that, for a given fixed integer $k \geq 1$, the algorithm described in Section 4 achieves the fraction $(\frac{k}{k+2})\mathcal{C}$ when it builds augmentations of size at most $k$. Broadly speaking, this result follows from two insights, stated below as propositions.

Ours is a randomized algorithm: even for fixed values of the link queue lengths, the new matching schedule is randomly generated. In a landmark paper [7], Tassiulas considered centralized randomized algorithms that can achieve the entire capacity region $\mathcal{C}$. The algorithms in [7] generate a random new matching in every time slot. Links are switched to be active according to the new matching generated at time $t$ if and only if the weight[6] of the new matching is greater than the weight of the previous matching used in time $t-1$.

Tassiulas showed that if a particular randomized algorithm has the ability to generate a *maximum-weight* matching for $q_t$ from $(q_t, I_{t-1})$ with probability at least $\delta > 0$, then that randomized algorithm achieves $\mathcal{C}$. The probability bound $\delta$ can depend on the graph $G$ but should to be independent of time $t$ and queue lengths $q_t$. So, for example, if the centralized algorithm picks uniformly from the set $\mathcal{I}$ of all matchings in $G$, then $\delta = \frac{1}{|\mathcal{I}|}$ and so the algorithm will achieve $\mathcal{C}$.

Even in the centralized case, comparing and merging two matchings takes $O(n)$ time in the worst case. Constant time algorithms can operate only in local neighborhoods, and it is unlikely that global comparison can be achieved in this case. Indeed, even generating a maximal matching[7] may be an $O(n)$ process in the worst case. In light of these realizations, it seems that constant-time algorithms may need to move away from global generation and global comparison. This may also imply a move away from trying to achieve the entire region $\mathcal{C}$ to achieving a fraction $\beta\mathcal{C}$.

Our first insight is that Tassiulas' result [7] extends to the case of approximately optimal matchings.

Proposition 1. *Given any $0 \leq \beta \leq 1$, suppose that an algorithm has a probability at least $\delta > 0$ of generating a matching with weight at least $\beta$ times the weight of the optimal. Then, $\beta\mathcal{C}$ can be achieved by switching links to the new matching when its weight is larger than the previous one, and the link weights are the associated queue lengths. The algorithm should generate the new matching $I_t$ from the old matching $I_{t-1}$ and current queue lengths $q_t$.*

The proof of Proposition 1 is presented in Appendix A. It is similar to the proof used in [7], but with a subtly modified Lyapunov function to take into account the $\beta$-approximation.

In light of Proposition 1, we look for ways to generate matchings of approximately optimal weight. Also, this generation should finish in constant time, implying that changes from $I_{t-1}$ to $I_t$ have to be small and local.

The following lemma provides a key step towards generating such local changes in $I_{t-1}$. Some quick notation: for any matching $I$, the term $Iq_t := \sum_{e \in I} q_t(e)$ refers to the sum of the queue lengths of links that are in $I$. Also, for any queue-length vector $q$, $I_q$ stands for the matching with the largest weight: $I_q q \geq Iq$ for all $I \in \mathcal{I}$.

Lemma 1. *Given any vector $q_t$ of queue lengths and existing matching $I_{t-1}$, there exists a set $\mathcal{A}^*$ of disjoint augmentations of $I_{t-1}$ such that*

$$(I_{t-1} \oplus \mathcal{A}^*)q_t \;\geq\; \left(\frac{k}{k+2}\right) I_{q_t} q_t$$

*and every augmentation in $A \in \mathcal{A}^*$ has size$(A) \leq k$.*

The above lemma states that there exists a set $\mathcal{A}^*$ of small (size at most $k$) disjoint augmentations such that the weight of the augmented matching will be close enough to the optimal weight. Given the existence of this $\mathcal{A}^*$, and in light of Proposition 1 above, all that is left to do is find an algorithm which provably finds $\mathcal{A}^*$ with probability at least $\delta$. We show that our algorithm indeed does this, and state the overall result below.

Proposition 2. *There exists a $\delta > 0$ such that our algorithm, operating using a fixed $k \geq 1$, generates a matching with weight within $\frac{k}{k+2}$ of the optimal with probability at least $\delta$, in any time slot $t$ and for any values of $(q_t, I_{t-1})$.*

The proof of Proposition 2 and related lemmas, including Lemma 1, is presented in Appendix B. Propositions 1 and 2 together imply that our algorithm achieves $\left(\frac{k}{k+2}\right)\mathcal{C}$.

## 6. SIMULATIONS

In this section, we investigate the performance of our class of algorithms via simulations. The primary purpose of this investigation is to look at the delay performance of our algorithms. In these simulations we compare our algorithm to the Maximal Matching (MM) algorithm, which is the following: at every time step a maximal matching[8] is generated and links in this matching are made active while others are made inactive. Note that this algorithm, at least as stated here and as simulated by us, is a centralized algorithm.

At present there do not exist no other constant-overhead algorithms that can guarantee arbitrary fractions of the stability region. So we would like to compare our algorithms to the existing constant-overhead algorithms [1, 2, 3]. Furthermore, each of these algorithms essentially tries to emulate Maximal Matching via enhanced contention resolution. So, for our simulations, we compare our algorithm directly with the Maximal Matching algorithm itself, with the implicit understanding that its performance would be better than that of the existing constant-time protocols.

Note of course that all of these algorithms, including the centralized Maximal Matching algorithm, can guarantee at best half of the capacity region, and so from the capacity viewpoint they are *not* comparable: our algorthms can capture any fraction $k/(k+2)$ of the capacity region for any $k \geq 1$. We emphasize that in this section we compare only the delay performance.

We ran our simulations on a simple grid network, shown in Figure 4. The network is an 11x11 grid with 121 nodes (represented by circles) and 220 links (represented by lines). Each link has the unit capacity, i.e. it can transmit one unit of data in one time slot, when active. The data arrival model is as follows: on each link, in each time-slot, one unit of data arrives with probability equal to the load on the link. Otherwise, no data arrives.

[6]Where the weights on the links at time $t$ are the queue lengths at time $t$.

[7]A maximal matching is one to which no link can be added.

[8]Recall that a *maximal* matching is any matching to which no link can be added without removing an existing link.

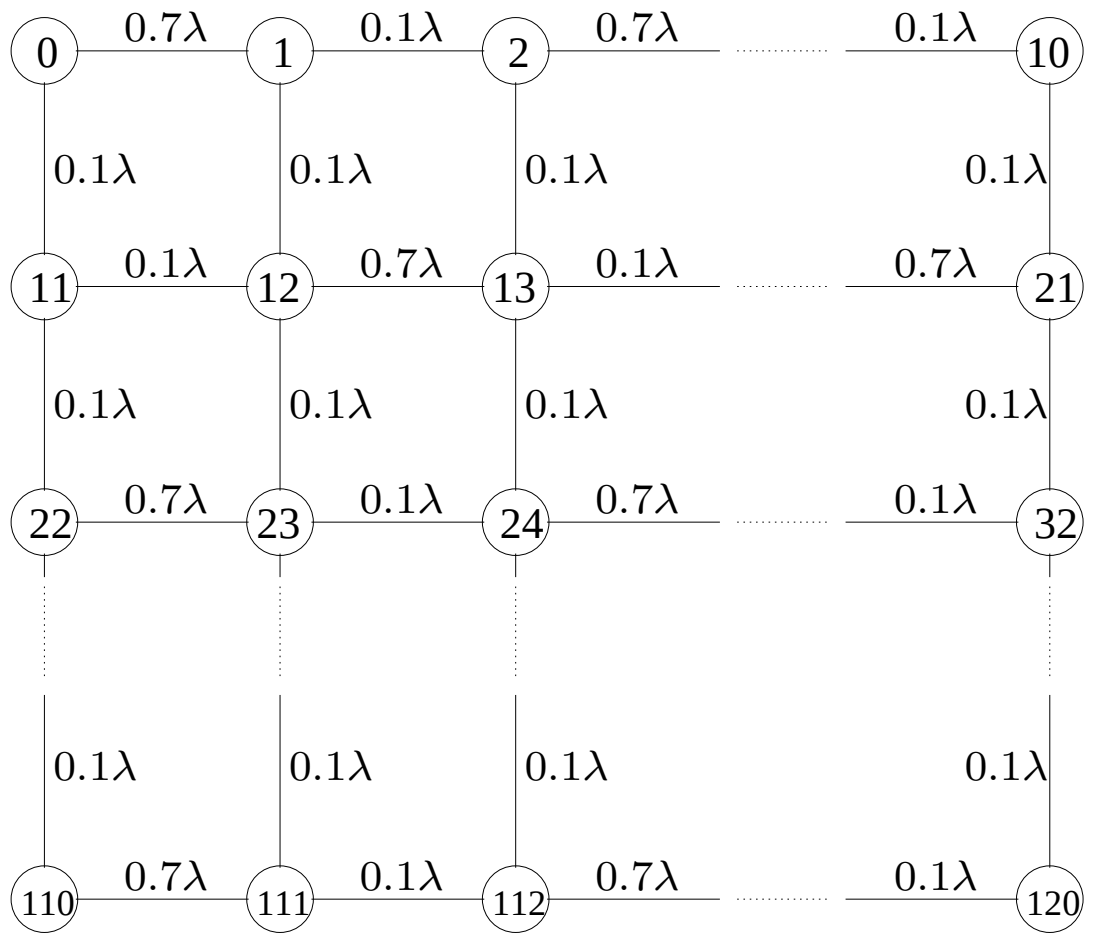

**Figure 4: Network Topology.**

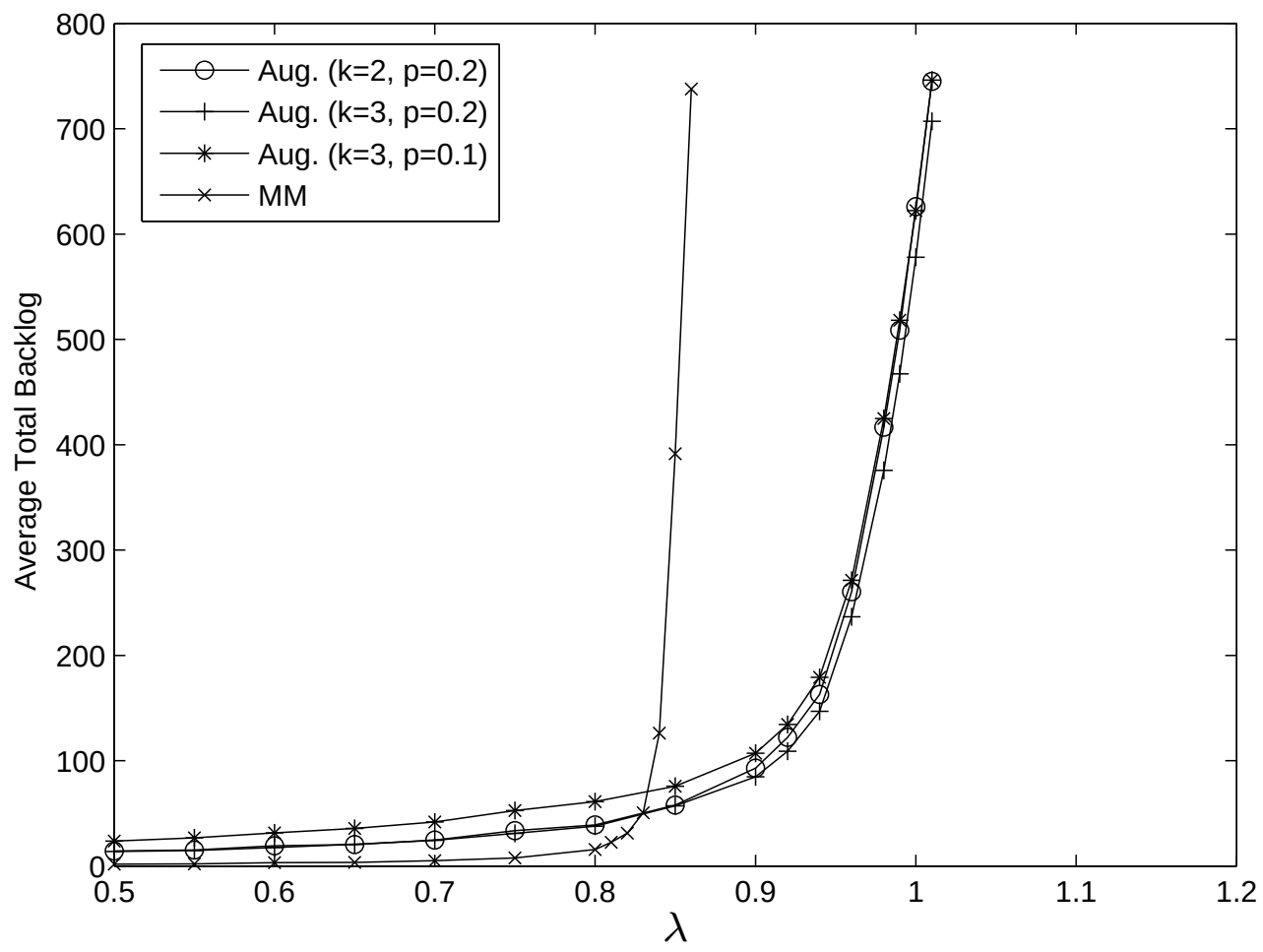

**Figure 5: Performance comparison to MM.**

Link loads in our example are parameterized by $\lambda$, and are either $0.1\lambda$ or $0.7\lambda$, as marked in the figure. It is clear that the capacity region of this network corresponds to $\lambda < 1$. The numbers 0.1 and 0.7 thus specify the *direction* in which we are investigating the capacity region, while $\lambda$ specifies the *extent* of the capacity region that is captured along that direction.

For any algorithm and network, as the load $\lambda$ increases the queue lengths will increase. Typically, as the load $\lambda$ grows towards a threshold the average of the node queue lengths starts increasing rapidly and beyond this $\lambda$ queues are unstable (i.e. grow with time). The threshold provides us with an estimate of the maximum load, i.e. the maximum portion of the capacity region, that the algorithm can handle with reasonable delays.

Notice that for our network, the maximum $\lambda$ sustainable by *any* algorithm – the "100% throughput load" – is equal to 1: loads higher than this will correspond to arrival rates that are outside the capacity region of the netwrok. This is easy to see from Figure 4: at any central node, the total load on all links incident on the node is $\lambda$. At most one of these links can be scheduled in one time slot, and thus a choice of $\lambda \geq 1$ will result in the queues being unstable.

In Figure 5, we plot the average total queue backlog of all links in the networks, in terms of $\lambda$. We have plotted the performance of our augmenting algorithm for the cases $\{k = 2, p = 0.2\}$, $\{k = 3, p = 0.2\}$, and $\{k = 3, p = 0.1\}$, as well as the performance of the MM algorithm.

Notice first of all that *even small-k implementations of our algorithm can stabilize higher loads than MM in reasonable networks like the one in the simulation.* In particular, while both MM and our algorithm for $k = 2$ both guarantee only half the capacity region, in the simulation our $k = 2$ algorithm achieves close to 100% thourghput while MM achieves only close to 85%. Also, the fact that $k = 2$ is already so close to 100% means that $k = 3$ performs similar to $k = 2$ in terms of achievable capacity. It however does have a slightly better delay performance.

We note that in our simulated network the actual performance of both MM and our algorithms is much better than what is implied by the respective performance guarantees, reflecting the fact that the guarantees represent contrived worst case scenarios for these algorithms. However, as our simulations also show, there is still a difference in the achievable throughput even over the enhanced performance of MM.

Also observe that for small loads our algorithm has higher queue sizes as compared to MM. This is because MM ensures that every link will be scheduled, or there will be an interfering link scheduled instead. When loads are low, the number of interfering links is small because the corresponding queues are empty for large portions of the time. This leads to good utilization. Ours algorithms are randomized, and do not ensure maximal usage. However, our simulations indicate that the queue lengths of our algorithms are not unreasonable.

Finally, we note that the parameter $p$ of a node becoming a seed is a free parameter that can be set by the designer. Our simluations seem to indicate that for reasonable values of $p$, the exact value does not seem to have an overwhelminng effect on performance. This seems to suggest that performance may not be too sensitive to the *exact* choice of the parameter.

To give another example, we ran another simulation on the same network but probed the performance in a different direction, i.e. with different relative loads on the links. In particular, we made the heavy horizontal links with $0.7\lambda$ into $0.89\lambda$, the lighter horizontal links $0.1\lambda$ are kept the same and the vertical links are made into $0.01\lambda$. That result is presented in Figure 6.

# 7. CONCLUSION AND DISCUSSION

This paper presents a novel class of simple distributed wireless scheduling algorithms. Our algorithms can achieve any fraction of the capacity region for data transmission, and require constant overheads that do not scale with network size. They represent the first instance in which both properties hold simultaneously. Our results are also interesting from a practical standpoint the due to simple structure of our algorithms, and our explicit accounting of the tradeoff between overhead and scheduling performance.

The objective of this paper is to design simple algorithms with appealing properties, not to announce a complete wireless protocol ready for implementation. In particular, careful studies need to be done to arrive at the value of $k$ that would give the best performance in practice. This will likely depend on the maximum length a scheduling cycle can have, given mobility and other aspects of the network that have been abstracted away in our model. An interesting avenue for possible algorithmic investigation is to see how our ideas adapt to designs for more general interference constraints.

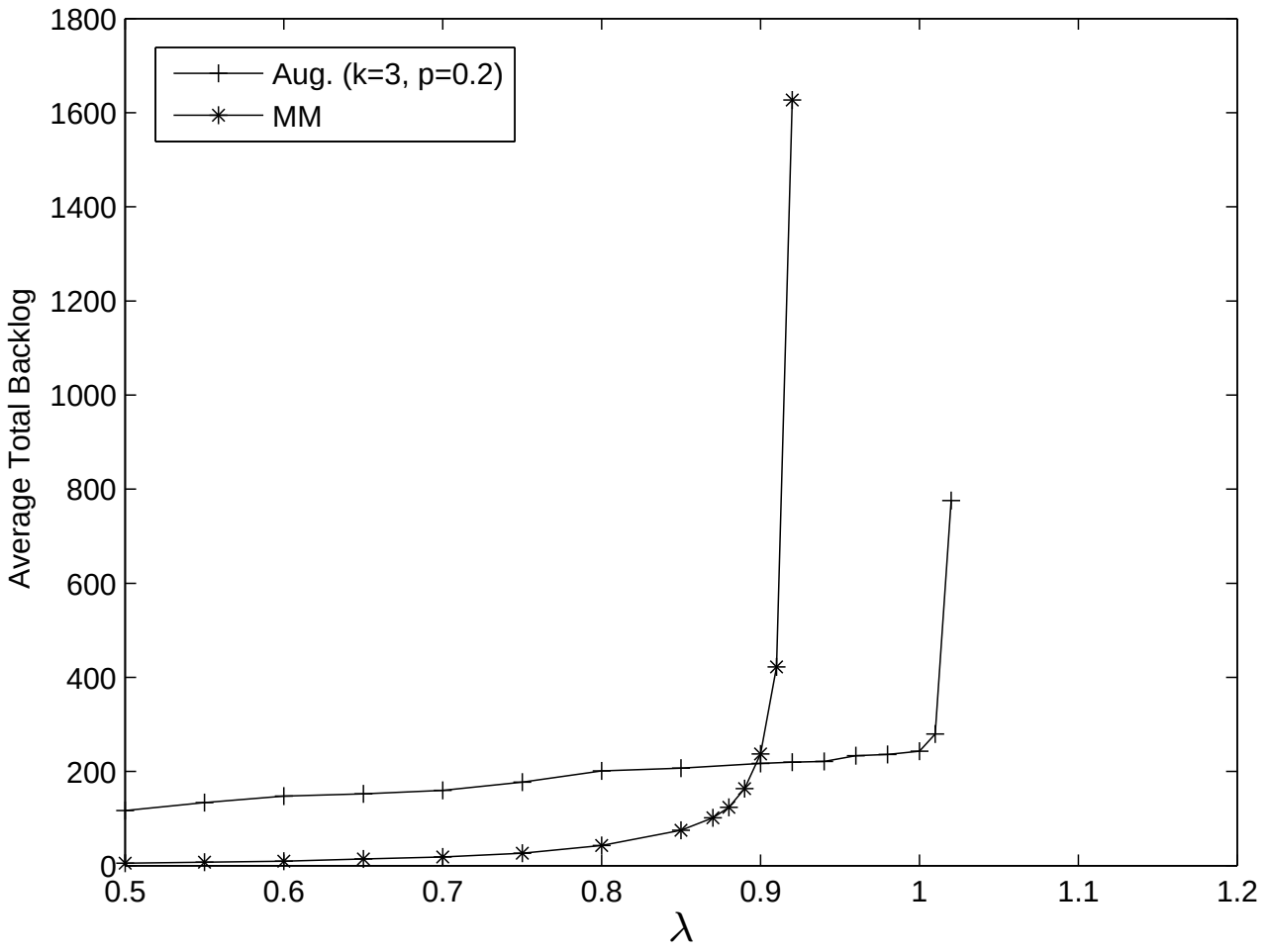

**Figure 6: Performance comparison to MM along a different direction.**

# APPENDIX

## A. PROPOSITION 1 AND RELATED LEMMAS

We now quickly develop some notation. In the following, we will multiply vectors with each other but drop the "transpose" symbol. Thus, for example, $q_t^2 := \sum_e (q_t(e))^2$ and $I_t q_t := \sum_e I_t(e) q_t(e)$. Also, for a given $q$ let $I_q$ denote any optimal matching for $q$, i.e. $I_q q \geq Iq$ for all $I \in \mathcal{I}$.

The proof below parallels the development in [7, Section 5], of which it is an extension. The difference in the proofs is that our Lyapunov function is a subtle modification of the one used in [7]. Indeed, when $\beta = 1$, our Lyapunov function is exactly the same as in [7].

Let $Y_t = (q_t, I_t)$, and let its (countable) state space be $\mathcal{Y}$. We are interested in algorithms for which the process $\{Y_t : t \geq 1\}$ is an irreducible and aperiodic Markov chain. This is clearly the case for our algorithm. To prove Proposition 1, we show that a certain Lyapunov function $V(Y)$, designed below, has negative drift outside of a fixed finite subset of $\mathcal{Y}$. The well-known Foster's criterion [20] will then imply the stability of the system.

For a given fixed $\beta$, we now define a Lyapunov function $V(Y) = V_1(Y) + V_2(Y)$ on the state space $(q, I)$, where

$$V_1(q, I) \; = \; q^2 \quad \text{and} \quad V_2(q, I) \; = \; [(\beta I_q q - Iq)_+]^2$$

In the above, we have used the convention $x_+ = \max\{x, 0\}$ for any real number $x$. In order to give the reader a roadmap of the proof, we first prove Proposition 1 directly, assuming certain lemmas hold. Subsequently we prove the lemmas.

**Proof of Proposition 1:**

The set $\{Y : V(Y) < B\}$ is a finite subset of the state space of $Y$. We will show that $V(Y_t)$ has negative drift outside this set, for $B$ large enough. From Lemmas 2 and 3 we have that if $a \in \beta\mathcal{C}$ then there exist positive constants $\epsilon, c_1, c_2$ and $c_3$ such that

$$\begin{aligned} E[V(Y_{t+1}) - V(Y_t)|Y_t] \;\; \leq \;\; & -\epsilon\sqrt{V_1(Y_t)} + (2 + c_2)\sqrt{V_2(Y_t)} \\ & - \delta V_2(Y_t) + c_1 + c_3 \end{aligned}$$

If $V(Y_t) \geq B$ then $V_1(Y_t) \geq (B - V_2(Y_t))_+$ and so the above equation becomes

$$\begin{aligned} E[V(Y_{t+1}) - V(Y_t)|Y_t] \;\; \leq \;\; & -\frac{\epsilon}{2}\sqrt{V_1(Y_t)} - \frac{\epsilon}{2}\sqrt{(B - V_2(Y_t))_+} \\ & + (2 + c_2)\sqrt{V_2(Y_t)} - \delta V_2(Y_t) \\ & + c_1 + c_3 \end{aligned}$$

Now, for $B$ large enough, it follows that

$$E[V(Y_{t+1}) - V(Y_t)|Y_t] \;\; \leq \;\; -\frac{\epsilon}{2}\sqrt{V_1(Y_t)} \quad \text{if } V(Y_t) \geq B$$

Given the above inequality, Foster's criterion implies the stability of the Markov chain $Y_t$, and thus of the queues $q_t$. See the proof of [7, Prop. 1] for the exact application of this result. Thus that the queues are stable whenever $a \in \beta\mathcal{C}$. ■

We now prove the lemmas used in the above proof.

Lemma 2. *For any $a \in \beta\mathcal{C}$ there exist constants $\epsilon$ and $c_1$ such that*

$$E[V_1(Y_{t+1}) - V_1(Y_t)|Y_t] \;\; \leq \;\; -\epsilon\sqrt{V_1(Y_t)} + 2\sqrt{V_2(Y_t)} + c_1$$

**Proof of Lemma 2:**

$$\begin{aligned} & E[V_1(Y_{t+1}) - V_1(Y_t)|Y_t] \\ & \quad = \; E[(q_{t+1} - q_t)(q_{t+1} + q_t)|Y_t] \\ & \quad = \; E[(A_{t+1} - I_t)(2q_t + A_{t+1} - I_t)|Y_t] \\ & \quad = \; E[(A_{t+1} - I_t)2q_t|Y_t] + E[(A_{t+1} - I_t)^2|Y_t] \qquad (3) \end{aligned}$$

We now look at each of the terms in (3) separately. For the first term,

$$\begin{aligned} & E[(A_{t+1} - I_t)2q_t|Y_t] \\ & \quad = \; (a - I_t)2q_t \\ & \quad = \; 2(a - \beta I_{q_t})q_t + 2(\beta I_{q_t} - I_t)q_t \qquad (4) \end{aligned}$$

Now, by the definition of $a$ in (1),

$$\begin{aligned} 2(a - \beta I_{q_t})q_t \;\; & = \;\; 2\left(\sum_m \lambda_m I^{(m)} q_t - \beta I_{q_t} q_t\right) \\ & \leq \;\; 2\left(\sum_m \lambda_m - \beta\right) I_{q_t} q_t \\ & \leq \;\; 2\left(\sum_m \lambda_m - \beta\right) \frac{1}{|E|}\sqrt{V_1(Y_t)} \end{aligned}$$

where the second inequality above follows from the definition of $I_{q_t}$ and the last inequality is proved as follows:

$$I_{q_t} q_t \;\; \geq \;\; \max_{e \in E} q_t(e) \;\; \geq \;\; \frac{1}{|E|}\sqrt{q_t^2} \;\; = \;\; \frac{1}{|E|}\sqrt{V_1(Y_t)}$$

Also, $2(\beta I_{q_t} - I_t)q_t \leq 2\sqrt{V_2(Y_t)}$ and so from (4) we have that

$$E[(A_{t+1} - I_t)2q_t|Y_t] \;\; \leq \;\; -\epsilon\sqrt{V_1(Y_t)} + 2\sqrt{V_2(Y_t)}$$

where $-\epsilon = 2\left(\sum_m \lambda_m - \beta\right)\frac{1}{|E|}$. Now,

$$E[(A_{t+1} - I_t)^2|Y_t] \;\leq\; E[(A_{t+1} + \mathbf{1})^2] \;\leq\; c_1$$

From the last two equations and (3) the lemma is proved. ■

Lemma 3. *If the algorithm generates a $\beta$-optimal matching with probability at least $\delta$, then there exist constants $c_2$ and $c_3$ such that*

$$E[V_2(Y_{t+1}) - V_2(Y_t)|Y_t] \leq -\delta V_2(Y_t) + c_2\sqrt{V_2(Y_t)} + c_3$$

**Proof of Lemma 3:**

Given $Y_t$, let $\mathcal{E}_\beta$ be the event $\{I_{t+1}q_{t+1} \geq \beta I_{q_{t+1}} q_{t+1}\}$. Then,

$$\begin{aligned}
E[V_2(Y_{t+1})|Y_t] &= P[\mathcal{E}_\beta|Y_t]\, E[V_2(Y_{t+1})|Y_t, \mathcal{E}_\beta] \\
&\quad + P[\mathcal{E}_\beta^c|Y_t]\, E[V_2(Y_{t+1})|Y_t, \mathcal{E}_\beta^c] \\
&\leq 0.P[\mathcal{E}_\beta|Y_t] \;+\; (1-\delta)E[V_2(Y_{t+1})|Y_t, \mathcal{E}_\beta^c] \\
&= (1-\delta)E[((\beta I_{q_{t+1}} - I_{t+1}) \\
&\qquad \times(q_t + A_{t+1} - I_t))^2|Y_t, \mathcal{E}_\beta^c] \qquad (5)
\end{aligned}$$

Now, by definition of $I_{q_t}$, we have that $\beta I_{q_{t+1}} q_t \leq \beta I_{q_t} q_t$ and $I_{t+1}q_t \geq\geq I_t q_t - |E|$. Thus,

$$\begin{aligned}
(\beta I_{q_{t+1}} - I_{t+1})\, q_t &\leq (\beta I_{q_t} - I_t)q_t + |E| \\
&\leq \sqrt{V_2(Y_t)} + |E|
\end{aligned}$$

Also,

$$(\beta I_{q_{t+1}} - I_{t+1})\,(A_{t+1} - I_t) \;\leq\; A_{t+1} + \mathbf{1}$$

Putting the above two equations into (5) proves the lemma, since $E[A_{t+1}^2] = M < \infty$. ■

# B. PROPOSITION 2 AND RELATED LEMMAS

Lemma 1 says there exists a "good set" $\mathcal{A}^*$ of augmentations: where each augmentation is not too large, and that augmenting $I_{t-1}$ using the set represents a certain amount of gain. We build up towards the proof of Lemma 1 by designing a candidate set $\mathcal{A}^*$ having augmentations of size at most $k$. We will then prove the gain it represents is as claimed by Lemma 1.

The *symmetric difference* $S := I_{t-1} \triangle I_{q_t}$ of matchings $I_{t-1}$ and $I_{q_t}$ is the set of links that are in exactly one of $I_{t-1}$ or $I_{q_t}$. Links that are in both or in neither are excluded from $S$. Consider the graph $G' := (V, S)$ containing only the links in $S$. Since each $I$ is a matching, a vertex in $G'$ has degree at most 2 and each connected component of $S$ is either an alternating path or even-length cycle. Also, each component is an augmentation of $I_{t-1}$, so we can define the size of a component in the same way as we defined the size of an augmentation of $I_{t-1}$.

For any component $C$, let $C_1 := C \cap I_{q_t}$ denote the links of $C$ in the optimal matching and $C_2 := C \cap I_{t-1}$ be the links in the current matching. Note that $C_1$ and $C_2$ are also matchings in $G$ and the terms $C_1 q_t$ and $C_2 q_t$ are as defined before for matchings. Also note that $size(C) = |C_1|$.

We build the set $\mathcal{A}^*$ from $S$ by finding a suitable set $\mathcal{A}_C$ in each component $C$ of $S$. The following two lemmas ensure this can be done.

Lemma 4. *If component $C$ of $S$ is a path then there exists a set $\mathcal{A}_C$ of disjoint augmentations contained in $C$ such that*

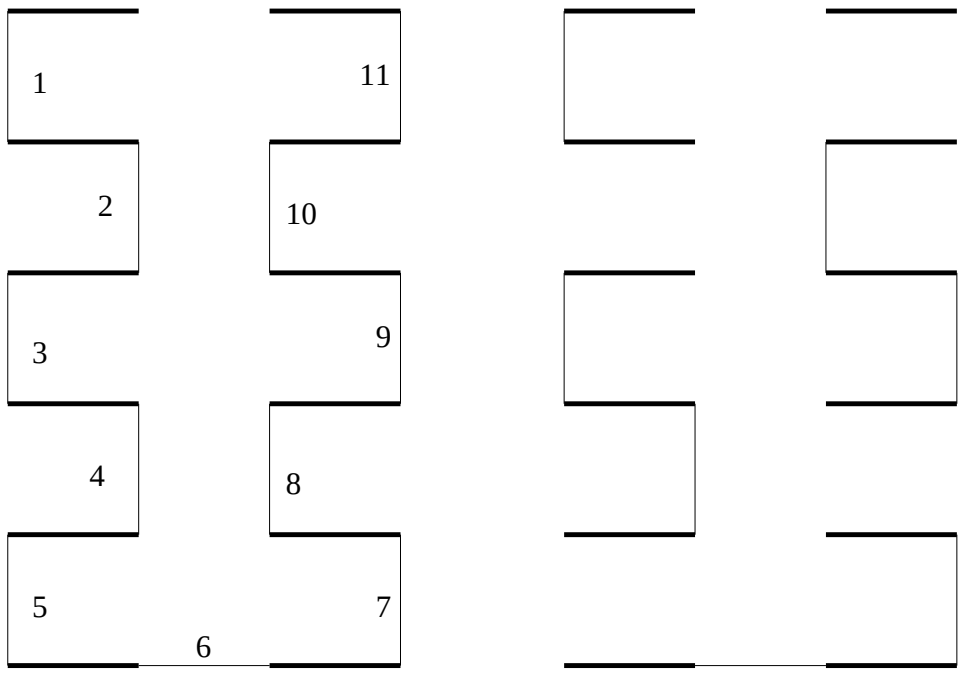

**Figure 7: The path on the left is a path component $C$ of $S$, with bold links denoting $C_1 = C \cap I_{t-1}$ and non-bold for $C_2 = I_{q_t}$. Links in $C_1$ are numbered, and $k = 2$. $C$ is cut up starting at link number 2, going anti-clockwise, to obtain the set $\mathcal{A}_2$ on the right. Note that $\mathcal{A}_2$ contains disjoint augmentations of size at most $k = 2$.**

1. *$size(A) \leq k$ for every $A \in \mathcal{A}_C$*
2. *$gain(\mathcal{A}_C) \geq \frac{k}{k+1} C_1 q_t - C_2 q_t$*

**Proof of Lemma 4:**

If $C$ is small, with $size(C) \leq k$, then let $\mathcal{A}_C = \{C\}$ be the set containing only $C$. This obviously satisfies both of the above conditions, since $gain(C) = C_1 q_t - C_2 q_t$. So in this case we are done.

So consider now a path $C$ with $size(C) \geq k+1$. Starting from any endpoint of the path, number all links in $C_1$. Let $e_1, \ldots, e_{k+1}$ be the first $k{+}1$ links in $C_1$. For each $e_i$, $1 \leq i \leq k+1$, build a set $\mathcal{A}_i$ of disjoint augmentations by deleting every $(k{+}1)^{th}$ link starting with $e_i$: i.e., link $e_m$ is deleted if and only if $m - i = 0$ or $m - i$ is divisible by $k+1$. Figure 7 shows this process for a simple example. After the deletions, each remaining fragment of $C$ will be an augmentation of $I_{t-1}$, and will have size at most $k$. These fragments together make the set $\mathcal{A}_i$ of disjoint augmentations.

Consider now the sets $\mathcal{A}_1, \ldots, \mathcal{A}_{k+1}$, made from the links $e_1, \ldots, e_{k+1}$ above, respectively. It is clear that each link in $C_2$ will be a member of all $k+1$ of these sets, and each link in $C_1$ will be a member of $k$ of the sets. Recall from (2) that the gain of any set $\mathcal{A}$ is the weight of all its links outside $I_{t-1}$ minus the weight of all its links in $I_{t-1}$. Thus,

$$\sum_{i=1}^{k+1} gain(\mathcal{A}_i) \;=\; kC_1 q_t - (k+1)C_2 q_t$$

which means that there exists at least one $j$ such that

$$gain(\mathcal{A}_j) \;\geq\; \frac{k}{k+1} C_1 q_t - C_2 q_t$$

Setting $\mathcal{A}_C = \mathcal{A}_j$ proves the lemma. ■

Lemma 4 shows the existence of a good set of disjoint augmentations in path components of $S$. We now use this to prove a slightly weaker result for cycle components of $S$.

Lemma 5. *If component $C$ of $S$ is a cycle, then there exists a set $\mathcal{A}_C$ of disjoint augmentations such that*

1. *every $A \in \mathcal{A}_C$ is contained in $C$, i.e. $A \subset C$, and has $size(A) \leq k$*

*2.* $gain(\mathcal{A}_C) \geq \frac{k}{k+2} C_1 q_t - C_2 q_t$

Note that the ratio is $\frac{k}{k+2}$ instead of $\frac{k}{k+2}$ as it was for paths.

**Proof of Lemma 5:**

If $size(C) \leq k$, set $\mathcal{A}_C = \{C\}$ and we are done. So now consider cycles $C$ in $S$ with $size(C) \geq k+1$.

Let $e \in C_1$ be the link with the smallest queue in $C_1$, (i.e. lowest weight): $q_t(e) \leq q_t(e')$ for all $e' \in C_1$. Consider path $\widehat{C} = C - e$, and define $\widehat{C}_1 := \widehat{C} \cap I_{q_t}$ and and $\widehat{C}_2 := \widehat{C} \cap I_{t-1}$ as before. Obviously, $C_2 = \widehat{C}_2$. Also, since $e$ was chosen to be the link in $C_1$ with smallest queue and $|C_1| = size(C) \geq k+1$ ,

$$\widehat{C}_1 q_t \;\geq\; \frac{|C_1|}{|C_1|+1} C_1 q_t \;\geq\; \frac{k+1}{k+2} C_1 q_t$$

Also, by Lemma 4, there exists a set $\mathcal{A}_{\widehat{C}}$ of disjoint augmentations of size at most $k$ in $\widehat{C}$ such that

$$\begin{aligned} gain(\mathcal{A}_{\widehat{C}}) &\geq \frac{k}{k+1} \widehat{C}_1 q_t - \widehat{C}_2 q_t \\ &\geq \frac{k}{k+1} \left( \frac{k+1}{k+2} C_1 q_t \right) - C_2 q_t \\ &= \frac{k}{k+2} C_1 q_t - C_2 q_t \end{aligned}$$

Setting $\mathcal{A}_C = \mathcal{A}_{\widehat{C}}$ proves the lemma. ■

We are now ready to build the set $\mathcal{A}^*$: for each component $C$ of $S$ add to $\mathcal{A}^*$ the augmentations in the corresponding $\mathcal{A}_C$ – where $\mathcal{A}_C$ is as specified by Lemmas 4 and 5 for paths and cycles respectively. We are now ready to prove Lemma 1.

**Proof of Lemma 1:**

Let $\mathcal{A}^*$ be as constructed above. Its gain will be the sum of the gains of each of the $\mathcal{A}_C$'s of which it is composed. So,

$$\begin{aligned} gain(\mathcal{A}^*) &= \sum_C gain(\mathcal{A}_C) \\ &\geq \frac{k}{k+2} \left( \sum_C C_1 q_t \right) - \left( \sum_C C_2 q_t \right) \\ &= \frac{k}{k+2} \left[ I_{q_t} q_t - (I_{q_t} \cap I_{t-1}) q_t \right] \\ &\qquad - \left[ I_{t-1} q_t - (I_{q_t} \cap I_{t-1}) q_t \right] \\ &\geq \frac{k}{k+2} I_{q_t} q_t - I_{t-1} q_t \end{aligned}$$

where $I_{q_t} \cap I_{t-1}$ is the matching consisting of links that are in both $I_{q_t}$ and $I_{t-1}$. The lemma's proof follows from the fact that

$$gain(\mathcal{A}^*) = (I_{t-1} \oplus \mathcal{A}^*) q_t - I_{t-1} q_t$$

by definition. ■

Now that we have shown the existence of a suitable set $\mathcal{A}^*$, all we need to do to prove Proposition 2 is to uniformly lower bound the probability of the algorithm generating $\mathcal{A}^*$.

**Proof of Proposition 2:**

Recall that in our algorithm a disjoint set of augmentations $\mathcal{A}$ is created and the ones with positive gain are switched to obtain $I_t$ from $I_{t-1}$. Thus,

$$I_t q_t \;\geq\; (I_{t-1} \oplus \mathcal{A}) q_t$$

So, it suffices to lower bound

$$P[\mathcal{A} = \mathcal{A}^* | q_t, I_{t-1}]$$

by a quantity that is independent of $(q_t, I_{t-1})$ but can depend on graph structure and $k$. Note of course that $\mathcal{A}^*$ is not independent of $(q_t, I_{t-1})$.

We will now provide a very naive lower bound to the above probability. Let $l = |\mathcal{A}^*|$ be the number of disjoint augmentations in $\mathcal{A}^*$. Choose one node in each of these augmentations as follows: if the augmentation is a path, choose one of its endpoints, and if it is a cycle choose any node. Let $\mathcal{E}$ be the event that *all* of the following are true:

- the algorithm generates $\mathcal{A} = \mathcal{A}^*$,
- the only seeds active in phase 1 are the nodes chosen above, and
- each augmentation's intended length (chosen at its seed) is equal to the actual length of that augmentation in $\mathcal{A}^*$ (i.e., no augmentation is a result of a cutoff due to contention).

Clearly,

$$P[\mathcal{A} = \mathcal{A}^* | q_t, I_{t-1}] \;\geq\; P[\mathcal{E} | q_t, I_{t-1}]$$

We now lower bound the right hand side. The probability that the nodes turn active or remain inactive as specified by $\mathcal{E}$ is $p^l (1-p)^{n-l}$. Further, the probability that the intended lengths are exactly as chosen is $k^{-l}$, since they are chosen uniformly from $1, \ldots, k$. Finally, we need each of the random link-addition choices (step 2 of the algorithm) of each of the above $l$ augmentations to exactly parallel the corresponding ones in $\mathcal{A}^*$. There are a total of at most $n$ such choices to be made, and the probability of each being the correct one is at least $\frac{1}{\Delta}$, where $\Delta$ is the maximum degree of the graph. Thus the probability of correct link choices is at least $\Delta^{-n}$.

Putting everything together, we have that

$$\begin{aligned} P[\mathcal{E} | q_t, I_{t-1}] &\geq p^l (1-p)^{n-l} k^{-l} \Delta^{-n} \\ &\geq \min \left\{ 1, \left( \frac{p}{1-p} \right)^n \right\} \left( \frac{1-p}{k\Delta} \right)^n \end{aligned}$$

The right hand side of the last inequality is now independent of $(q_t, I_{t-1})$, providing us with the required uniform lower bound. Setting $\delta = \min \left\{ 1, \left( \frac{p}{1-p} \right)^n \right\} \left( \frac{1-p}{k\Delta} \right)^n$ completes the proof of the proposition. ■

## C. REFERENCES

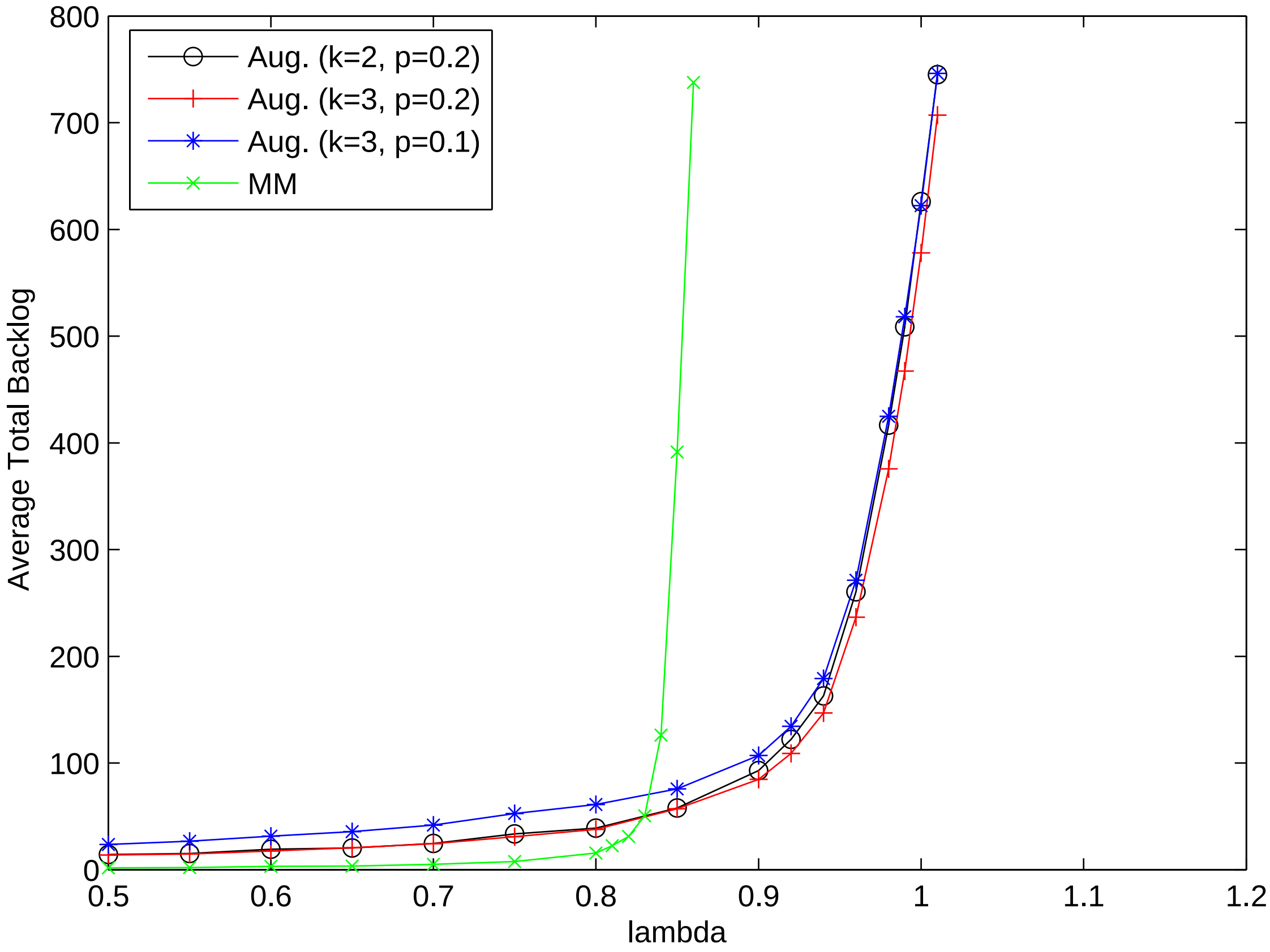